\documentclass[letter]{aa} 

\usepackage{graphicx}
\usepackage{txfonts}
\usepackage{xcolor}
\usepackage{multirow}
\usepackage{multicol}
\usepackage{array}
\usepackage[normalem]{ulem}

\newcommand{\feh}{$\langle \rm{[Fe/H]} \rangle$}
\newcommand{\ali}{A(Li)}
\newcommand{\medt}{$\overline{t}$}
\newcommand{\sfehtot}{$\sigma(\rm{[Fe/H]})_{\rm tot}$}
\newcommand{\sfehtop}{$\sigma(\rm{[Fe/H]})_{\rm top6}$}
\newcommand{\smgfetot}{$\sigma(\rm{[Mg/Fe]})_{\rm tot}$}
\newcommand{\smgfetop}{$\sigma(\rm{[Mg/Fe]})_{\rm top6}$}

\newcommand\Tstrut{\rule{0pt}{2.6ex}}         

\usepackage[allcolors=blue]{hyperref}

\makeatletter
\renewcommand*\aa@pageof{, page \thepage{} of \pageref*{LastPage}}
\makeatother

\begin{document}

    \title{The \emph{Gaia}-ESO Survey: Probing the lithium abundances in old  metal-rich dwarf stars in the solar vicinity}


    \author{M.~L.~L.~Dantas\inst{\ref{camk}}
        \and
        G.~Guiglion\inst{\ref{mpia}, \ref{aip}}
        \and
        R.~Smiljanic\inst{\ref{camk}}
        \and
        D.~Romano\inst{\ref{inaf_bologna}}
        \and
        L.~Magrini\inst{\ref{inaf_oaa}}
        \and
        T.~Bensby\inst{\ref{lundobs}}
        \and
        C.~Chiappini\inst{\ref{aip}}
        \and
        E.~Franciosini\inst{\ref{inaf_oaa}}
        \and
        S.~Nepal\inst{\ref{aip},\ref{ipa}}
        \and
        G.~Tautvai{\v s}ien{\. e}\inst{\ref{lithuania}}
        \and
        G.~Gilmore\inst{\ref{cambridge}}        
        \and 
        S.~Randich\inst{\ref{inaf_oaa}}         
        \and
        A.~C.~Lanzafame\inst{\ref{catania}}     
        \and
        U.~Heiter \inst{\ref{uppsala_univ}}     
        \and
        L. Morbidelli\inst{\ref{inaf_oaa}}      
        \and
        L.~Prisinzano\inst{\ref{inaf_palermo}}  
        \and
        S.~Zaggia\inst{\ref{inaf_padova}}       
    }

    \institute{Nicolaus Copernicus Astronomical Center, Polish Academy of Sciences, ul. Bartycka 18, 00-716, Warsaw, Poland\\ \email{mlldantas@protonmail.com or mlldantas@camk.edu.pl} \label{camk}
    \and
    Max Planck Institute for Astronomy, K\"onnigstuhl 17, 69117, Heidelberg, Germany \label{mpia}
    \and
    Leibniz-Institut f\"ur Astrophysik Potsdam (AIP), An der Sternwarte 16, 14482 Potsdam, Germany \label{aip}
    \and
    INAF -- Osservatorio di Astrofisica e Scienza dello Spazio di Bologna, Via Gobetti 93/3, 40129 Bologna, Italy \label{inaf_bologna}
    \and
    INAF -- Osservatorio Astrofisico di Arcetri, Largo E. Fermi, 5, 50125 Firenze, Italy \label{inaf_oaa}
    \and
    Lund Observatory, Department of Astronomy and Theoretical Physics, Box 43, SE-221 00 Lund, Sweden \label{lundobs}
    \and
    Institut für Physik und Astronomie, Universität Potsdam, Karl-Liebknecht-Str. 24/25, 14476 Potsdam, Germany \label{ipa}
    \and 
    Institute of Theoretical Physics and Astronomy, Vilnius University, Saul\.{e}tekio av. 3, LT-10257 Vilnius, Lithuania \label{lithuania}
    \and
    Institute of Astronomy, University of Cambridge, Madingley Road, Cambridge CB3 0HA, United Kingdom \label{cambridge}
    \and
    Dipartimento di Fisica e Astronomia, Sezione Astrofisica, Universit\'{a} di Catania, via S. Sofia 78, 95123, Catania, Italy\label{catania}
    \and 
    Observational Astrophysics, Department of Physics and Astronomy, Uppsala University, Box 516, 75120 Uppsala, Sweden \label{uppsala_univ}
    \and
    INAF -- Osservatorio Astronomico di Palermo, Piazza del Parlamento 1, 90134, Palermo, Italy\label{inaf_palermo}
    \and
    INAF -- Padova Observatory, Vicolo dell'Osservatorio 5, 35122 Padova, Italy \label{inaf_padova}
    }

    \date{Received XXX; accepted XXX}

 
  \abstract
   {Lithium (Li) is a fragile element that is produced in a variety of sites but can also be very easily depleted in stellar photospheres. Radial migration has been reported to explain the decrease in the upper envelope of Li measurements observed for relatively old metal-rich dwarf stars in some surveys.}
   {We test a scenario in which radial migration could affect the Li abundance pattern of dwarf stars in the solar neighbourhood. This may confirm that the Li abundances in these stars cannot serve as a probe for the Li abundance in the interstellar medium (ISM). In other words, to probe the evolution of the Li abundance in the local ISM, it is crucial that stellar intruders be identified and removed from the adopted sample.}
   {We used the high-quality data (including Li abundances) from the sixth internal Data Release of the \emph{Gaia}-ESO survey. In this sample we grouped stars by similarity in chemical abundances via hierarchical clustering. Our analysis treats both measured Li abundances and upper limits.}
   {The Li envelope of the previously identified radially migrated stars is well below the benchmark meteoritic value (<3.26 dex); the star with the highest detected abundance has A(Li) = 2.76 dex. This confirms the previous trends observed for old dwarf stars (median ages $\sim$ 8 Gyr), where Li decreases for [Fe/H]$\gtrsim$0.}
   {This result is supporting evidence that the abundance of Li measured in the upper envelope of old dwarf stars should not be considered a proxy for the ISM Li. Our scenario also indicates that the stellar yields for [M/H]>0 should not be decreased, as recently proposed in the literature. Our study backs recent studies that claim that old dwarfs on the hot side of the dip are efficient probes of the ISM abundance of Li, provided atomic diffusion does not significantly lower the initial Li abundance in the atmospheres of metal-rich objects.}


   \maketitle

\section{Introduction}

The chemical abundances estimated from stellar atmospheres can reveal many details about the origin of stars, including the composition of their original gas clouds. However, a few elements, such as lithium (Li), are fragile and subject to changes because of stellar evolution \citep[see][for recent overviews on this topic]{Smiljanic2020, randich2021}.

For over three decades, the upper envelope of the Li abundance distribution of warm dwarf stars (i.e. $T_\mathrm{eff} \gtrsim 5700$ K) has been customarily used as a good tracer of the evolution of the Li abundance in the interstellar medium (ISM) of the Milky Way \citep[MW; see][]{Rebolo1988}. The trend of increasing Li abundance with increasing metallicity ([Fe/H] or [M/H]), from the Spite plateau value \citep[interpreted as the primordial Li abundance; see][]{SpiteSpite1982} to the higher meteoritic value, was thence attributed to the Li pollution from both stars and cosmic rays \citep[e.g.][]{Romano1999}. Lately, several studies have reported an unexpected decrease in this envelope for [M/H]>0 \citep[i.e. the super-solar metallicity regime;][]{DelgadoMena2015, guiglion_2016, Bensby2018, fu_2018}. 

\citet{Guiglion2019} suggested that such a decrease is due to old dwarfs that migrated from the inner regions of the MW disc while depleting their photospheric Li during their travel to the solar neighbourhood. One implication of such a scenario is that the upper envelope of Li is, therefore, not a robust tracer of the ISM Li abundance. This has recently been confirmed by \citet{charbonnel_2021} using dwarfs observed by the GALactic Archaeology with HERMES \citep[GALAH;][]{DeSilva2015} survey. These authors selected dwarfs on the hot side of the Li dip that preserved their original Li due to a very narrow convective zone \citep[see, for instance,][Sect. 4 therein]{Smiljanic2010}. Using open clusters and field stars from the \emph{Gaia}-ESO Survey sixth internal Data Release (GES-iDR6), \citet[][]{romano2021} also showed that young (ages $<$ 1--2 Gyr) dwarf and pre-main-sequence stars do not show any decrease in the upper envelope of Li abundances in the metal-rich regime \citep[see also][]{randich2020}. Additionally,  \citet[][]{Bensby2020} analysed a sample of microlensed dwarfs of the bulge and reported a decrease in the abundance of Li in the old (ages $>$ 8 Gyr) metal-rich regime.

In this Letter, using GES-iDR6 \citep{Gilmore2012, Gilmore2022, Randich2013, Randich_2022}  \ion{Li}{i} abundances [\ali], we demonstrate the validity of the \citet{Guiglion2019} scenario. To that end, we use a sample of old super-metal-rich stars currently inhabiting the solar vicinity that seem to have migrated from the inner regions of the MW. We also discuss the differences and similarities for two cases of Li measurements (detections and upper limits) and verify the effects of age. This Letter is structured as follows: in Sect. \ref{sec:data_methodology} we present the data and methodology applied; in Sect. \ref{sec:results_discussion} we present and discuss our results; and in Sect. \ref{sec:conclusions} we present the conclusions and final remarks.

\section{Data and methodology} \label{sec:data_methodology}

In this study we made use of the large dataset of dwarf stars described in \citet{Dantas2022}, selected from over 1400 targets observed with the Ultraviolet and Visual Echelle Spectrograph \citep[UVES;][]{Dekker2000} available in GES-iDR6. Their atmospheric parameters were computed with the codes described in \citet{Smiljanic2014} and combined with the Bayesian methodology described in Worley et al. (in prep.; see also a summary in \citealt{Gilmore2022}). The Li abundances were derived as described in \citet{Franciosini2022}.

We ordered the stars in a sequence of chemical enrichment steps and separated groups of stars that have similar abundances. We did that in a space of 21 dimensions. This includes 18 elements, three of which  are represented by both neutral and ionised species; they are as follows: \ion{C}{i}, \ion{Na}{i}, \ion{Mg}{i}, \ion{Al}{i}, \ion{Si}{i}, \ion{Si}{ii}, \ion{Ca}{i}, \ion{Sc}{ii}, \ion{Ti}{i}, \ion{Ti}{ii}, \ion{V}{i}, \ion{Cr}{i}, \ion{Cr}{ii}, \ion{Mn}{i}, Fe,\footnote{The abundance of Fe was estimated via the [Fe/H] provided by GES-iDR6. Hence, there is no ionisation level associated with this abundance.} \ion{Co}{i}, \ion{Ni}{i}, \ion{Cu}{i}, \ion{Zn}{i}, \ion{Y}{ii}, \ion{and Ba}{ii}. This analysis was done using hierarchical clustering \citep[HC; e.g.][]{Murtagh&Contreras2012, Murtagh2014}. We stress that \ion{Li}{i} was not used in this classification. In Level 1 of this separation, we have 6 main groups; 11 subgroups in Level 2; and 48 subgroups in Level 3 \citep[for more details, we refer the reader to Fig. 1 and Table 2 in][]{Dantas2022}. 

It is worth mentioning that we tested the non-local thermodynamic equilibrium effects for \ali~by using the prescription of \citet[][]{Wang2021}. For the super-solar sample (described in Sect. \ref{subsec:smr_sample}, which includes the most-metal-rich  sample), the median corrections are at 0.04 and 0.05 depending on the subgroup, with a standard deviation of 0.01 in all cases. Therefore, we did not apply these corrections to our sample, as they are too small. In any case, we display the corrections in Table \ref{tab:super_solar_nlte_li_corr}.

We use, in the current discussion, four subgroups with super-solar metallicity ([Fe/H]$\gtrsim$0) identified in Level 2 of the HC classification. These subgroups are numbered 10, 1, 2, and 3 (in order of increasing \feh). For consistency, we adopted the same numbering of groups and subgroups as in \citet[][]{Dantas2022}. Each subgroup was cleaned for missing \ion{Li}{i}  data, and no constraints in terms of signal-to-noise ratio or uncertainty requirements were applied to the \ali; these groups have respectively 192, 220, 78, and 91 stars.\footnote{All the stars in each group with Li measurements. The groups have, in total, 194, 221, 78, and 93 stars.} Subgroups 2 and 3 compose the        most-metal-rich (MMR) sample of stars (i.e. [Fe/H] $\geq$ 0.15) explored in detail by \citet{Dantas2022}.

The ages used here were estimated using \textsc{unidam} \citep[][]{Mints2017, Mints2018}; \textsc{unidam} is a Bayesian code that uses PARSEC isochrones \citep[][]{Bressan2012} to estimate stellar ages by fitting photometric and spectroscopic data from each star. Typical errors are usually around 1--2 Gyr \citep[see][]{Mints2017, Mints2018}.

\section{Results and discussion} \label{sec:results_discussion}

In \citet{Dantas2022} we analysed a sample of super-metal-rich stars, with median ages (\medt) of $\sim$8 Gyr. These stars show orbital features that suggest that they moved from their original birthplace in the inner Galaxy to the solar vicinity. Dynamical processes that can relocate a star from the inner regions of the MW to the solar vicinity are `blurring' and `churning' \citep[see e.g.][]{Schonrich2009a, Halle2015}. Blurring is the epicyclic movement of a star around the Galactic centre in eccentric orbits with no change in angular momentum; churning is the stellar movement characterised by the change in angular momentum due to the interaction between the star and non-axisymmetric structures (such as the bar and spiral arms). Churning is what is usually meant when the term radial migration is used. Indeed, the group of super-metal-rich stars that we discuss in this work seems to have mostly suffered from churning. The features that support this hypothesis include low orbital eccentricities, a large variation in vertical Galactic scale heights, and an incompatibility with radial metallicity gradient models \citep[see the full discussion in][]{Dantas2022}. 

We analysed the \ali~ in the stars, which migrated from the inner Galaxy, with the goal of probing the hypothesis proposed by \citet{Guiglion2019}. In their scenario, the decrease in the upper envelope of the \ali~observed for super-solar-metallicity dwarfs is due to an interplay between stellar evolution and radial migration. Stars formed in the inner Galaxy should exhibit higher \ali~when compared to their counterparts at solar radii. This is due to a higher star formation efficiency, which results in a higher abundance of all chemical elements, including Li (see Fig. 2 in \citealt{Guiglion2019}; Fig. 8 in \citealt{romano2021}; and \citealt{chiappini2009} for more details on the underlying chemical evolution model). However, stellar evolution acting while the stars are ageing on their path towards the solar neighbourhood leads to strong Li depletion in their atmospheres. They are, therefore, observed with sub-meteoritic \ali, creating an inversion in the observed \ali~upper envelope for [M/H]>0. 

We note here that in the discussion below two benchmark values for \ali~are used: the Li content of meteorites (\ali=3.26 dex, which is indicative of the ISM Li abundance at the Sun's birth; \citealt{lodders2009}), and the current photospheric solar abundance (\ali=1.05 dex, \citealt{Grevesse2007}).

In the following subsections, we discuss the results for the super-solar sample ([Fe/H]$\gtrsim$0) found in the HC clustering (Sect. \ref{subsec:entire_sample}) and for the MMR subgroups (Sect. \ref{subsec:smr_sample}).

\subsection{Super-solar sample} \label{subsec:entire_sample}

\begin{figure*}
    \centering
    \includegraphics[width=0.49\linewidth]{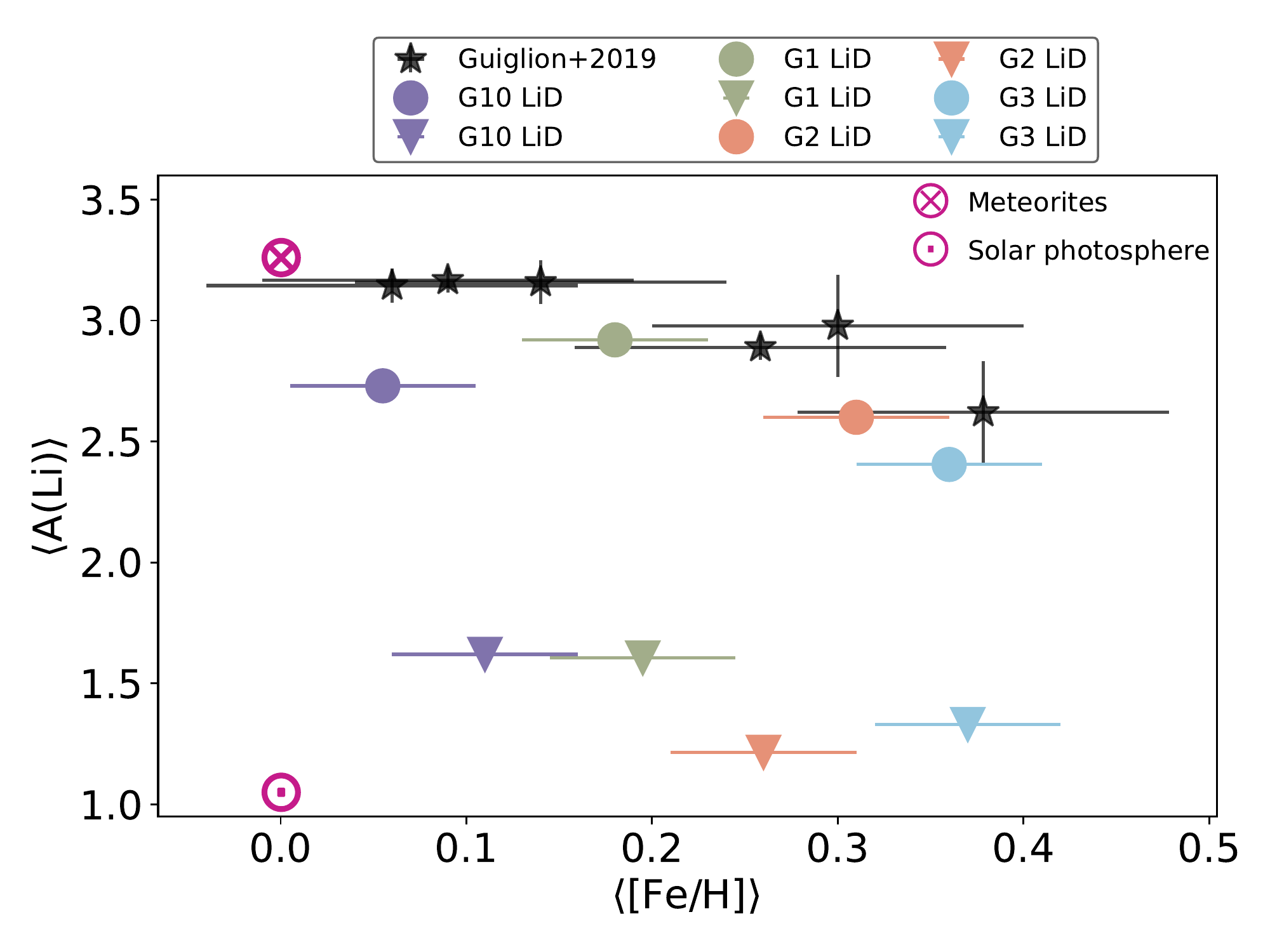}
    \includegraphics[width=0.49\linewidth]{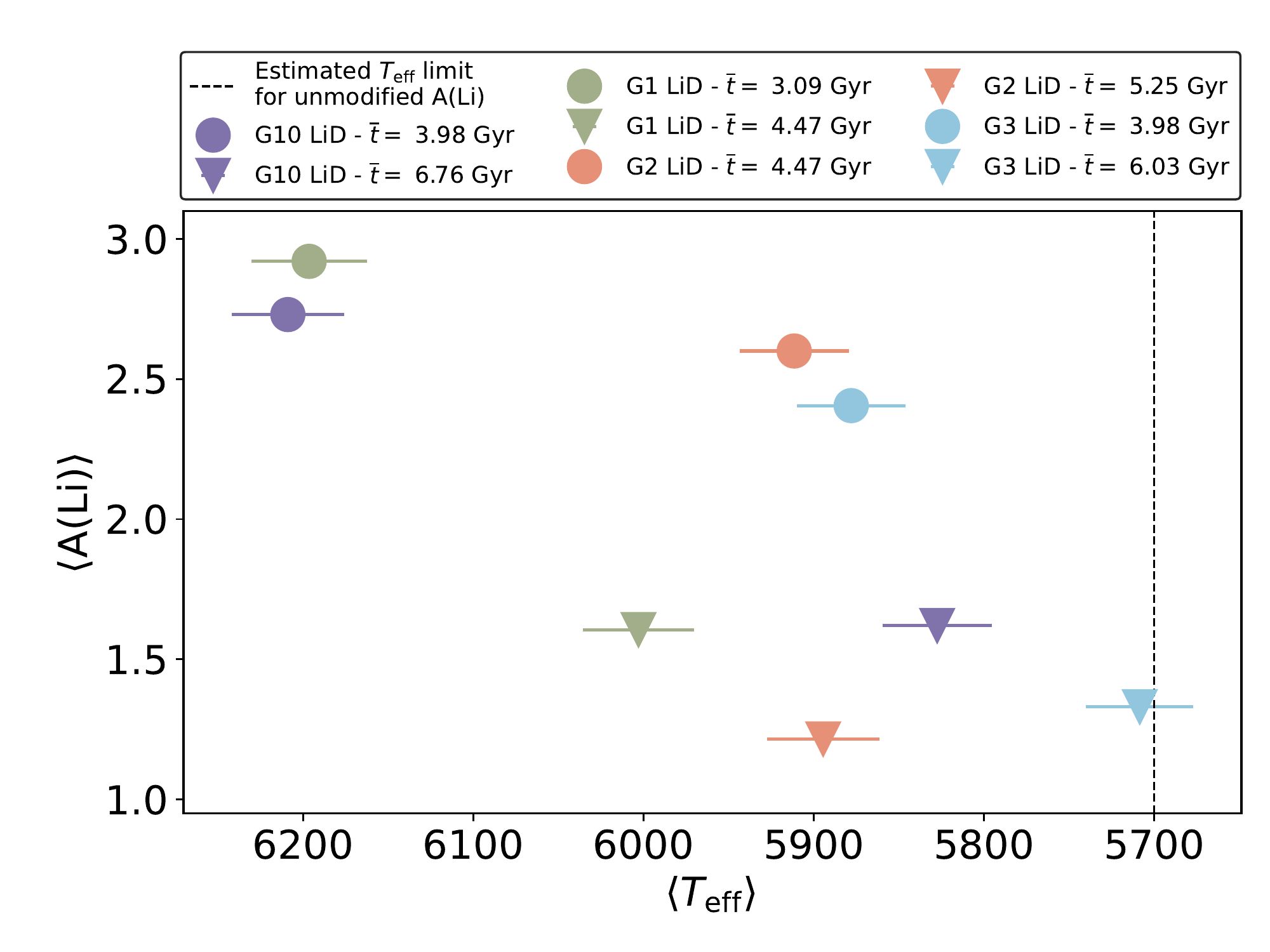}
     \caption{\ali~vs \feh~and $\langle T_{\rm eff} \rangle$ respectively for the super-solar groups. \emph{Left panel:} $\langle {\rm A(Li)} \rangle$ vs \feh~for the super-solar groups of the sample, split into those with a direct detection of Li (LiD -- round markers) and those with an upper limit estimate (LiUL -- inverted triangle markers). In this representation, only the median of the top six stars with the highest \ali~is shown in each marker. Dark magenta markers have been added to represent the meteoritic ({\tiny{$\bigotimes$}}) and solar photospheric ({\tiny $\bigodot$}) values. The median ages of the stars represented by each marker are displayed on the legend. The black star-shaped markers display the data from \citet{guiglion_2016}. It is worth mentioning that the errors associated with \ali~are very small, and that is why they are not seen. \emph{Right panel:} $\langle {\rm A(Li)} \rangle$ vs $\langle T_{\rm eff} \rangle$. It is possible to see that \ali~seems to decrease with decreasing $T_{\rm eff}$. Indeed, the warmer temperatures seem to have a protective effect on \ali~due to their thinner convective layers. A straight dotted black line is added to depict an approximate estimation of $T_{\rm{eff}}$ (i.e. at $\sim$5700K) for unmodified \ali,~as shown in \citet[][]{romano2021}.}
    \label{fig:li_feh_top6}
\end{figure*}

The left panel of Fig. \ref{fig:li_feh_top6} depicts the median values of \ali~versus [Fe/H] ($\langle {\rm A(Li)} \rangle$ x \feh) for the four MMR subgroups (Level 2 division for the HC described in \citealt{Dantas2022}). We split the sample in two based on the type of Li measurements: detected (LiD) and upper limit (LiUL).

It is noteworthy that the behaviour seen in the data from \citet{guiglion_2016} is very similar to that seen in the \citet{Dantas2022} star sample: \ali~declines for increasing [Fe/H] at super-solar values (\feh $\gtrsim$0). This similarity is especially important for LiD stars. It is worth mentioning that these subgroups are quite chemically homogeneous; their standard deviation for [Fe/H] and [Mg/Fe] is typically around one order of magnitude lower than their median values. The reader can assess their homogeneity using Table \ref{tab:supersolarsigmas}.

The right panel of Fig. \ref{fig:li_feh_top6} is the same as the left panel but with $\langle T_{\rm eff} \rangle$ instead of \feh. The right panel helps us understand the potential cause for this depletion. Indeed, \ali~increases with increasing $T_{\rm eff}$ as warmer stars tend to deplete less Li because of their thinner convective regions. It is also worth noting that there is an age difference between LiD and LiUL stars (\medt~is displayed on the label of Fig. \ref{fig:li_feh_top6}). The values of \medt~seem to be systematically higher for LiUL stars compared to their LiD counterparts. This decrease in Li with increasing age could be due to its destruction in the interior of these stars as they age; it could also be an effect of $T_{\rm eff}$, as we will discuss further in Sect. \ref{subsec:smr_sample}.

\subsection{Super-metal-rich subsample (MMR subgroups)} \label{subsec:smr_sample}

\begin{figure*}
    \centering
    \includegraphics[width=\linewidth]{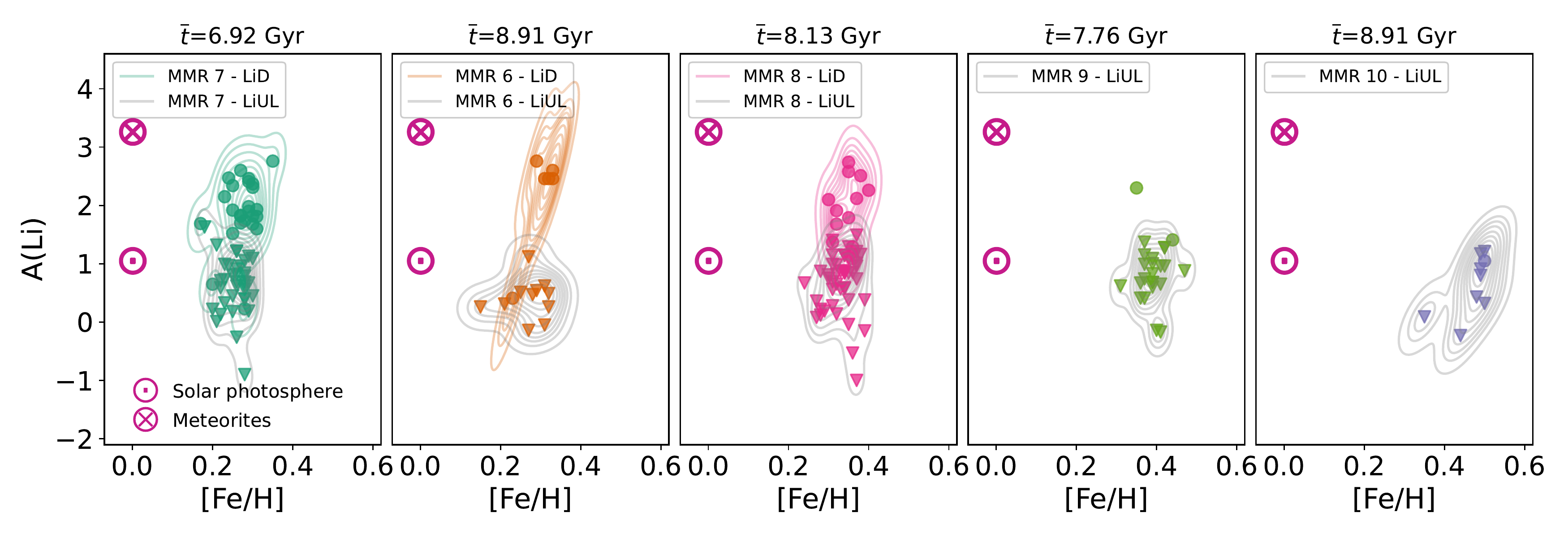}
    \caption{\ali~vs [Fe/H] in the shape of a scatter plot and a 2D-Gaussian kernel density plot of the MMR subgroups (from 6 to 10) in order of increasing \feh. Round and inverted-triangle markers depict, respectively, LiD and LiUL measurements. The 2D-kernel densities are either in the same colour as the scatter markers (for LiD) or in grey (for LiUL). 2D-kernel densities are not shown for subgroups 9 and 10 in the case of LiD due to the low number of stars with detected Li in each of these subgroups (two and one, respectively). As in Fig. \ref{fig:li_feh_top6}, additional markers were added to depict both the meteoritic and solar photospheric \ali. The MMR group colours are the same as in \citet{Dantas2022}. The errors are omitted to ease the visualisation. The median ages of each MMR subgroup are depicted at the top of each subplot; it is worth noting that subgroup 10 (purple) has a smaller $\overline{t}$ than depicted in \citet[][]{Dantas2022}. The reason for this difference is due to the removal of stars with missing Li measurements.}
    \label{fig:mmr_feh_li}
\end{figure*}

\begin{table*}
    \centering
    \caption{Median values for a few selected parameters of the MMR subgroups.}
    \begin{tabular}{l|l|l|l|l}
        MMR subgroup & \medt & \feh  & $\langle {\rm A(Li)} \rangle$ LiD & $\langle {\rm A(Li)} \rangle$ LiUL \\
                      & (Gyr) & (dex) & (dex) & (dex)\\
        \hline
        \hline
         7 (dark green)  & 6.92 & 0.27 & 1.90 & 0.69 \\
         6 (orange)      & 8.91 & 0.29 & 2.46 & 0.47 \\
         8 (pink)        & 8.13 & 0.34 & 2.00 & 0.82 \\
         9 (light green) & 7.76 & 0.39 & 1.86 & 0.78 \\
        10 (purple)      & 8.91 & 0.49 & 1.05 & 0.62 \\
         
    \end{tabular}
    \label{tab:mmr_general}
    \tablefoot{Columns list ages (\medt), metallicities (\feh), and Li abundances ($\langle {\rm{A}}_{Li} \rangle$) for both detected (LiD) and upper limit (LiUL) measurements. The subgroups are listed in order of increasing \feh. We note that the median age for subgroup 10 (purple) is smaller than described in \citet[][]{Dantas2022}. This difference is due to the stars removed with missing Li in the current paper.}
\end{table*}

Figure \ref{fig:mmr_feh_li} shows the individual stars in each MMR subgroup seen in Fig. \ref{fig:li_feh_top6} (which is represented by groups 2 and 3 in the Level 2 division of the HC), as well as their respective 2D-Gaussian kernel densities. These stars are the MMR sample from \citet{Dantas2022}, with the same colours as those therein. Additionally, Table \ref{tab:mmr_general} shows the general characteristics of the MMR subgroups (\feh, $\langle {\rm{A(Li)}} \rangle$, and \medt). It seems that the older the group, the farther the Li abundances of the stars get from the meteoritic value.

As discussed in \citet{Dantas2022}, these different subgroups seem to have migrated from different radii in the inner Galaxy. The stars in subgroup 10 are especially interesting, as they seem to have migrated from the innermost regions of the MW; they have the highest \medt~(along with subgroup 6\footnote{Subgroup 10 has a slightly lower age than presented in \citet[][]{Dantas2022}, and that is due to the removal of a few stars with missing Li measurements in the current paper.}) and are the ones most depleted in Li as well. It is noticeable that MMR subgroup 9 is the second youngest subgroup of our sample, but its members have very low \ali. This is probably due to the effects of radial migration, since this group migrated from the one of the innermost regions of the MW (second only to subgroup 10). This reinforces the role of radial migration in Li depletion in old metal-rich stars. The upper Li envelope of radially migrated stars then shows a decrease, as reported in \citet{Guiglion2019}, while the recent literature reported a constant Li ISM abundance in the super-solar-metallicity regime when younger stars are included in the sample \citep{randich2020, charbonnel_2021, romano2021}.

\begin{figure*}
    \centering
    \includegraphics[width=\linewidth]{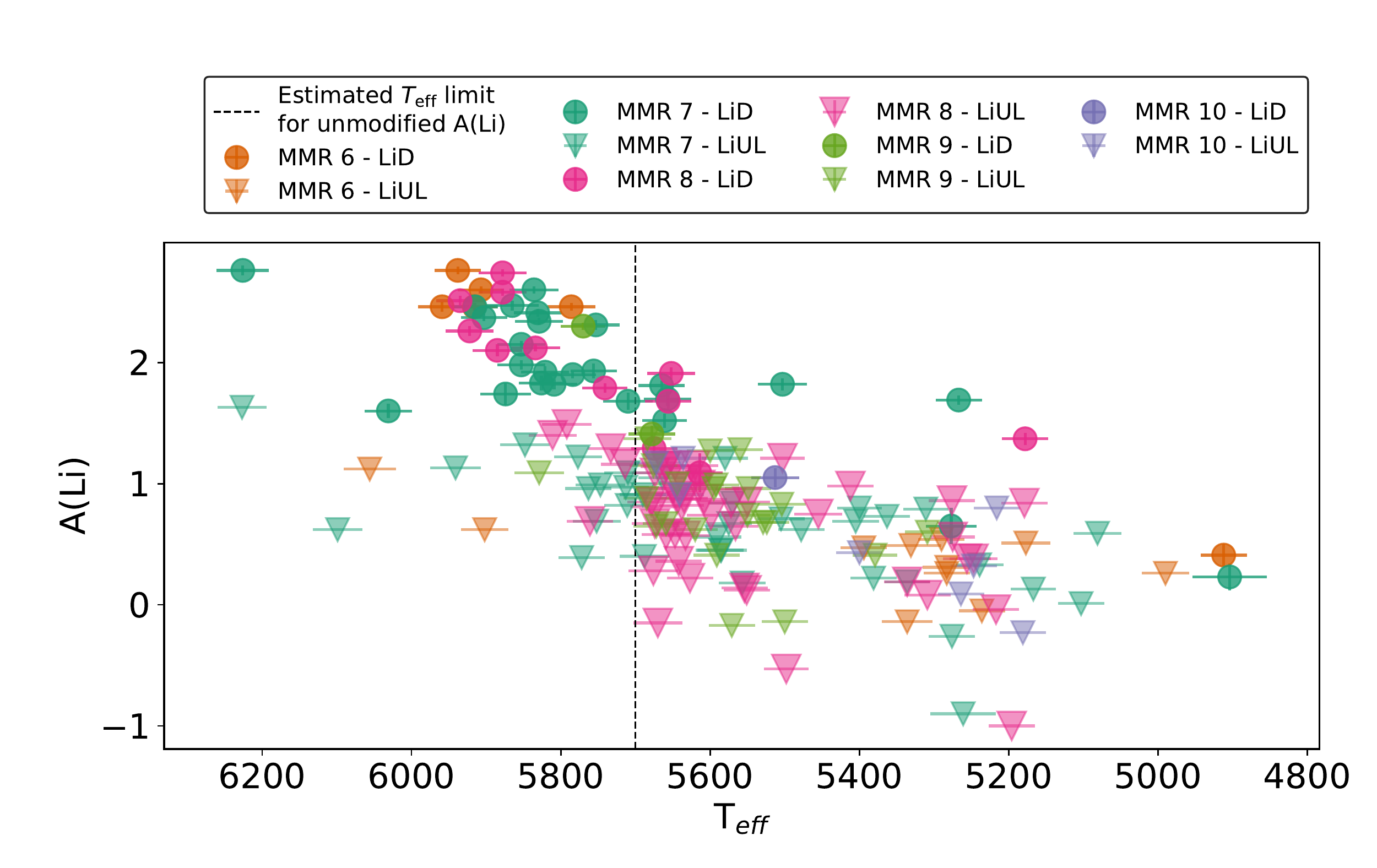}
    \caption{\ali~vs $T_{\rm{eff}}$ for all MMR stars. Stars with LiD are represented by the round markers, whereas those with LiUL are displayed using inverted triangles. The colours that differentiate each MMR subgroup are the same as in Fig. \ref{fig:mmr_feh_li}. A straight dotted black line is added to depict an approximate estimation of $T_{\rm{eff}}$ (i.e. at $\sim$5700K) for unmodified \ali,~as shown in \citet[][]{romano2021}.}
    \label{fig:mmr_li_teff}
\end{figure*}

In Fig. \ref{fig:mmr_li_teff} we display \ali~against the $T_{\rm{eff}}$ for all the stars in the MMR subgroups. It is noticeable that the temperatures of these stars are not high enough to have their initial \ali~fully preserved. This image supports the idea that old and cooler dwarf stars cannot be used as proxies for the ISM \ali. By contrast, the samples analysed in \citet[][]{randich2020} and \citet[][]{romano2021} also include some young stars (ages $<$ 1-2 Gyr), which prevented the detection of an \ali~decrease in the metal-rich regime. Additionally, according to the analysis performed by \citet[][see Fig. 4 therein]{romano2021}, there is a strong Li depletion at temperatures around 5700K, which is due to the effects of the stellar convective envelope. It is noteworthy that the stars in subgroup 10  in this work (see its main features in the above paragraph) do not reach temperatures above 5672K (hottest star therein). Therefore, its Li depletion could be caused by both stellar evolution and radial migration. For more details on these stars (such as the Kiel diagram), the reader is referred to \citet[][]{Dantas2022}.

In summary, the works of \citet[][]{randich2020} and \citet[][]{romano2021} explored the values for \ali~in metal-rich stars with younger ages, in the field and in open clusters, which can naturally have higher \ali. On the other hand, in \citet{Dantas2022} we identified a sample of old super-metal-rich stars that depict several dynamical features that indicate they were formed in the inner Galaxy. This sample is the best to use to probe the hypothesis proposed by \citet[][]{Guiglion2019}. Indeed, this scenario, in which the decrease in the upper envelope of Li abundances is caused by an interplay between radial migration and stellar evolution, is supported by the Li pattern of our (likely) migrated MMR stars. This is especially the case for the stars in subgroup 10.

\section{Conclusions} \label{sec:conclusions}

In this Letter we have probed the scenario described by \citet{Guiglion2019}, by checking whether old radially migrating dwarfs are responsible for the observed decrease in the Li envelope for [M/H]$\gtrsim$0 seen in some stellar samples. We used the sample of stars available in GES-iDR6 that was previously analysed by \citet{Dantas2022}. The super-metal-rich stars ([Fe/H] $\geq$ 0.15 dex) in this sample show several chemo-dynamical characteristics expected for stars that have migrated from the inner regions of the Galaxy, such as low eccentricities and incompatibility with radial-metallicity gradient models. The upper envelope of the Li abundance for these groups of migrators indeed decreases with increasing [Fe/H]. The implications of this validation are diverse:

\begin{enumerate}
    \item MMR subgroup 10 is the subgroup that has, overall, the largest Li depletion among the MMR subgroups. It is also one of the oldest subgroups (along with subgroup 6) and the one that seems to have migrated from the innermost regions of the Galaxy.
    
    \item Considering the validity of the \citet{Guiglion2019} scenario, there is no need for reducing the stellar yields of red-giant branch stars at super-solar metallicities, which was a possibility explored by \citet{prantzos_2017}.
    
    \item Consequently, the upper envelope of Li as traced by old metal-rich stars is not representative of the true Li content of the ISM.
    
    \item The age and the effective temperature play a role in the depletion of Li, as seen in both the super-solar and the MMR samples. Stars with an upper limit for \ali~seem to be systematically older compared to those with detected Li. 
    
    \item The most solid tracers of Li evolution in the ISM appear to be younger main-sequence stars on the hot side of the Li dip \citep{randich2020, romano2021}; yet, the effects of atomic diffusion on the surface composition of the MMR dwarf stars still need to be fully assessed \citep{charbonnel_2021}.    

\end{enumerate}

In the near future, the 4-metre Multi-Object Spectroscopic Telescope \citep[4MOST;][]{2019Msngr.175....3D} and the \textit{William Herschel }Telescope Enhanced Area Velocity Explorer \citep[WEAVE;][]{10.1117/12.2312031} will provide Li abundances for millions of stars, allowing us to put more constraints on its puzzling chemical evolution. Additionally, these surveys will allow us to find more examples of stars that may have migrated from the inner Galaxy to the solar neighbourhood.

\begin{acknowledgements}
M.~L.~L.~Dantas and R. Smiljanic acknowledge support by the National Science Centre, Poland, project 2019/34/E/ST9/00133. TB was supported by grant No. 2018-04857 from the Swedish Research Council. The authors thank the anonymous referee for the comments on the manuscript. M.~L.~L.~Dantas also thanks Miuchinha for the love and support. This work made use of the following on-line platforms: \texttt{slack}\footnote{\url{https://slack.com/}}, \texttt{github}\footnote{\url{https://github.com/}}, and \texttt{overleaf}\footnote{\url{https://www.overleaf.com/}}. This work was made with the use of the following \textsc{python} packages (not previously mentioned): \textsc{matplotlib} \citep{Hunter2007}, \textsc{numpy} \citep{Harris2020}, \textsc{pandas} \citep{mckinney-proc-scipy-2010}, \textsc{seaborn} \citep{Waskom2021}. This work also benefited from \textsc{topcat} \citep{Taylor2005}. All figures were made with a qualitative palette from \url{https://colorbrewer2.org} -- credits: Cynthia Brewer, Mark Harrower, and The Pennsylvania State University. Based on data products from observations made with ESO Telescopes at the La Silla Paranal Observatory under programme ID 188.B-3002. These data products have been processed by the Cambridge Astronomy Survey Unit (CASU) at the Institute of Astronomy, University of Cambridge, and by the FLAMES/UVES reduction team at INAF/Osservatorio Astrofisico di Arcetri. These data have been obtained from the \textit{Gaia}-ESO Survey Data Archive, prepared and hosted by the Wide Field Astronomy Unit, Institute for Astronomy, University of Edinburgh, which is funded by the UK Science and Technology Facilities Council. This work was partly supported by the European Union FP7 programme through ERC grant number 320360 and by the Leverhulme Trust through grant RPG-2012-541. We acknowledge the support from INAF and Ministero dell' Istruzione, dell' Universit\`a' e della Ricerca (MIUR) in the form of the grant "Premiale VLT 2012". The results presented here benefit from discussions held during the \textit{Gaia}-ESO workshops and conferences supported by the ESF (European Science Foundation) through the GREAT Research Network Programme. This publication makes use of data products from the Wide-field Infrared Survey Explorer, which is a joint project of the University of California, Los Angeles, and the Jet Propulsion Laboratory/California Institute of Technology, funded by the National Aeronautics and Space Administration. This work has made use of data from the European Space Agency (ESA) mission {\it Gaia} (\url{https://www.cosmos.esa.int/gaia}), processed by the {\it Gaia} Data Processing and Analysis Consortium (DPAC, \url{https://www.cosmos.esa.int/web/gaia/dpac/consortium}). Funding for the DPAC has been provided by national institutions, in particular the institutions participating in the {\it Gaia} Multilateral Agreement.
\end{acknowledgements}

\bibliographystyle{aa}        
\bibliography{paper}      

\appendix

\section{Additional material}

In Table \ref{tab:super_solar_nlte_li_corr} we display the summary statistics for the estimated non-local thermodynamic equilibrium Li corrections stratified by super-solar group. It is worth noting that the super-solar sample includes the MMR sample. The table shows that the corrections would have been very small, and therefore, we decided not to use them in our analysis. These corrections were performed by using the code described in \citet[][]{Wang2021}.

\begin{table*}
    \centering
    \caption{Summary statistics for the Li NLTE corrections.}
    \begin{tabular}{l|c|c|c|c|c|c|c}
    \multicolumn{8}{c}{Lithium NLTE corrections for our super-solar sample.}\\
    \hline
     Level 2 subgroup &  mean & $\sigma$ &   min &   25\% &   50\% &   75\% &   max \\
                      & (dex) & (dex)    & (dex) &  (dex) & (dex)  & (dex)  & (dex) \\
     \hline
     \hline
     G01 (green)      & -0.05 &  0.01 & -0.06 & -0.05 & -0.05 & -0.05 & -0.01 \\
     G02 (orange)     & -0.04 &  0.01 & -0.06 & -0.05 & -0.05 & -0.04 & -0.01 \\
     G03 (blue)       & -0.04 &  0.01 & -0.06 & -0.04 & -0.04 & -0.04 & -0.02 \\
     G10 (purple)     & -0.05 &  0.01 & -0.07 & -0.06 & -0.06 & -0.05 &  0.00 
    \end{tabular}
    \tablefoot{The summary statistics for the Li NLTE corrections for the super-solar sample are displayed above. The table includes the mean, standard deviation ($\sigma$), minimum correction, 25\textsuperscript{th} percentile, 50\textsuperscript{th} percentile, 75\textsuperscript{th} percentile, and maximum correction. These values were not applied to the Li abundances, but they show that such corrections would have been marginal for our stars.}

    \label{tab:super_solar_nlte_li_corr}
\end{table*}

In Table \ref{tab:supersolarsigmas} we display the standard deviation ($\sigma$) for the metallicity parameters ([Fe/H] and [Mg/Fe]) for the super-solar groups in order of increasing \feh: G10, G01, G02, and G03 (G02 and G03 being the MMR group). It is possible to find \sfehtot, \sfehtop, \smgfetot, and \smgfetop, respectively, therein, where `tot' refers to the total sample and `top6' to the top six Li-rich stars.

\begin{table*}
    \centering
    \caption{Standard deviation ($\sigma$) for [Fe/H] and [Mg/Fe] for all the groups with super-solar metallicity.}
    \begin{tabular}{l|l|c|c|c|c}
        Level 2 subgroup     & Li detection & \sfehtot & \sfehtop & \smgfetot & \smgfetop \\
        \hline
        \hline
        \multirow{2}{*}{G10 (purple)} & LiD  & 0.044 & 0.034 & 0.049 & 0.049 \Tstrut \\
                                      & LiUL & 0.047 & 0.076 & 0.064 & 0.047  \\
        \hline
        \multirow{2}{*}{G01 (green)}  & LiD  & 0.047 & 0.034 & 0.072 & 0.085 \Tstrut \\
                                      & LiUL & 0.050 & 0.019 & 0.067 & 0.028 \\
        \hline
        \multirow{2}{*}{G02 (orange)} & LiD  & 0.038 & 0.038 & 0.058 & 0.054 \Tstrut \\
                                      & LiUL & 0.036 & 0.038 & 0.064 & 0.061 \\
        \hline
        \multirow{2}{*}{G03 (blue)}   & LiD  & 0.050 & 0.019 & 0.043 & 0.038 \Tstrut \\
                                      & LiUL & 0.056 & 0.039 & 0.057 & 0.039 \\
                                      
 \end{tabular}
    \label{tab:supersolarsigmas}
    \tablefoot{The values are stratified by the type of Li detection and given both for the entire sample within the four groups and for the stars with the top 6 highest Li detection. The groups are listed in order of increasing $\langle \rm{[Fe/H]} \rangle$. The Table is useful in order to evaluate the level of chemical homogeneity of the super-solar groups  (Level 2 of the HC).}
\end{table*}

\end{document}